\documentclass[cits]{PoS}

\title{Polarisation of mJy radio sources}

\ShortTitle{Polarisation of mJy radio sources}

\author{\speaker{J. M. Stil} \\
        Centre for Radio Astronomy\\
        The University of Calgary\\
        E-mail: \email{stil@ras.ucalgary.ca}
       }

\author{A. R. Taylor\\
        Centre for Radio Astronomy\\
        The University of Calgary\\
        E-mail: \email{russ@ras.ucalgary.ca}
        }

\author{M. Krause\\
        Max Planck Intitut f\"ur Radioastronomie, Bonn, Germany\\
        E-mail: \email{mkrause@mpifr-bonn.mpg.de}
        }

\author{R. Beck\\
        Max Planck Intitut f\"ur Radioastronomie, Bonn, Germany\\
        E-mail: \email{rbeck@mpifr-bonn.mpg.de}
        }

\abstract{ Predictions of the number of faint polarised radio sources
that can be detected by SKA pathfinder telescopes and the SKA depend
on the polarisation properties of radio sources with a total flux
density around 1 mJy. Total intensity source counts suggest a
transition in the dominant population from AGN to galaxies around this
flux density, and the properties of brighter radio sources may not be
representative for this fainter population. We show that unresolved
spiral galaxies can be highly polarised radio sources, up to
$\sim$20\% polarised at 4.8 GHz. This result is partly based on
observations of nearby galaxies, including galaxies with significant
deviations from axial symmetry and other peculiarities.  A first
analysis of polarised source counts divided into steep-spectrum AGN,
flat-spectrum AGN and star forming galaxies is presented, including a
prediction of polarised source counts to $\mu$Jy levels.}

\FullConference{From Planets to Dark Energy: The Modern Radio Universe\\
		 1$^{st}$-5$^{th}$ October 2007\\
		 The University of Manchester, United Kingdom}

\begin{document}

\section{Science with polarised extragalactic sources}

SKA pathfinder projects currently under development aim to do $\mu$Jy
sensitivity wide-area spectro-polarimetric surveys. These surveys will
be capable of detecting many thousands of polarised sources, but the
angular resolution $\lesssim 1'$ will not be sufficient to resolve
these sources. Nevertheless, many aspects of the SKA key science goal
of the origin and evolution of cosmic magnetism can be studied with
these deep polarisation surveys. By increasing the number density of
polarised sources with observed rotation measure (the density of the
rotation measure grid \cite{bg04}), the magneto-ionic medium of the
Galaxy and in nearby galaxies \cite{stepanov2007} can be mapped in
much more detail than before. Figure~1 shows an example from
\cite{stil2007} of what such studies may reveal.  The left panel shows
the number density of polarised sources in the NVSS \cite{condon1998}
in the region of the Gum nebula. A strong deficiency below the mean
background of 6.6 polarised sources per square degree is visible in a
ring centered on $(l,b)=(-106^\circ,+4^\circ)$ that coincides with an
arc of H$\alpha$ emission associated with the Gum nebula shown in
Figure~1 (b). The deficiency in background polarised sources is the
result of depolarisation resulting from a high rotation measure of
material in a shell.

A polarimetric survey with ASKAP \cite{johnston2007} may provide
$\sim$ 20 rotation measures per square degree over most of the
sky. This number comes from a model-dependent extrapolation of the
observed polarised source counts, $N(p) = 1.25 p^{-2.5}$ for polarised
flux density $20 > p > 0.7$ mJy \cite{tucci2004,taylor2007}. The
question how faint the slope of the polarised source counts remains
constant is crucial for the extrapolation to $\mu$Jy levels. Knowledge
of the polarisation of radio sources of a few mJy total flux density
is required for a realistic extrapolation. The first such
extrapolation \cite{bg04} was based on the polarisation properties of
sources brighter than 80 mJy. These authors already noted that such an
extrapolation would be a lower limit to the number of polarised
sources if the fractional polarisation $\Pi_0$ of faint sources is
higher, as found by a number of authors
\cite{tucci2004,taylor2007,mesa2002,sadler2006}. Moreover, radio
source counts suggest a population change around 1 mJy, so it is not
clear that bright radio sources are representative for faint radio
sources.

\begin{figure}
\centerline{\resizebox{6.0cm}{!}{\includegraphics[angle=0]{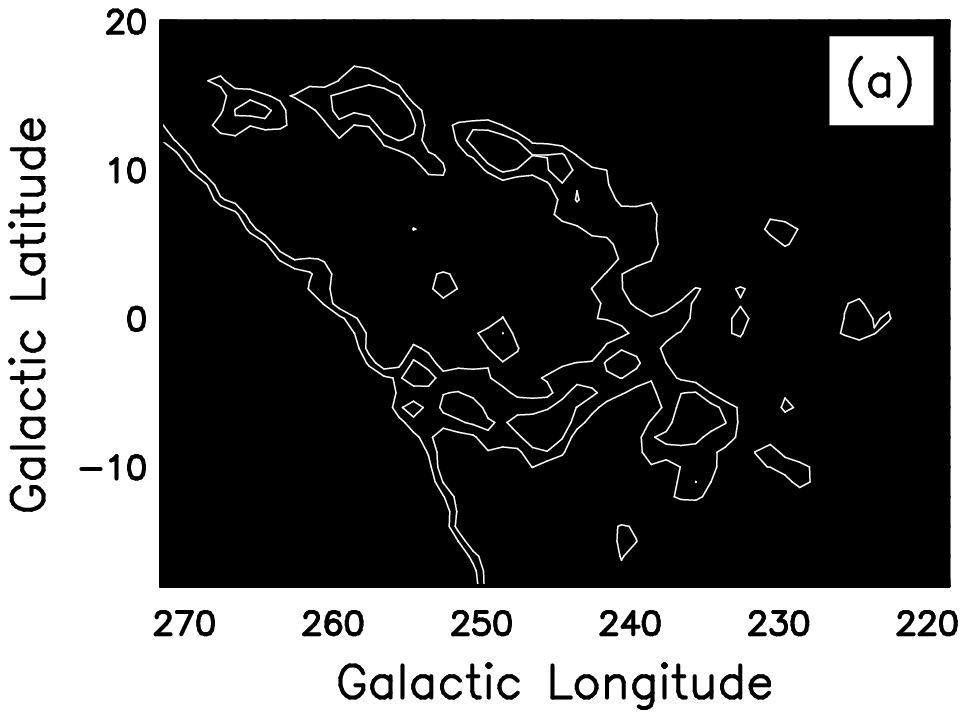}}
\resizebox{6.0cm}{!}{\includegraphics[angle=0]{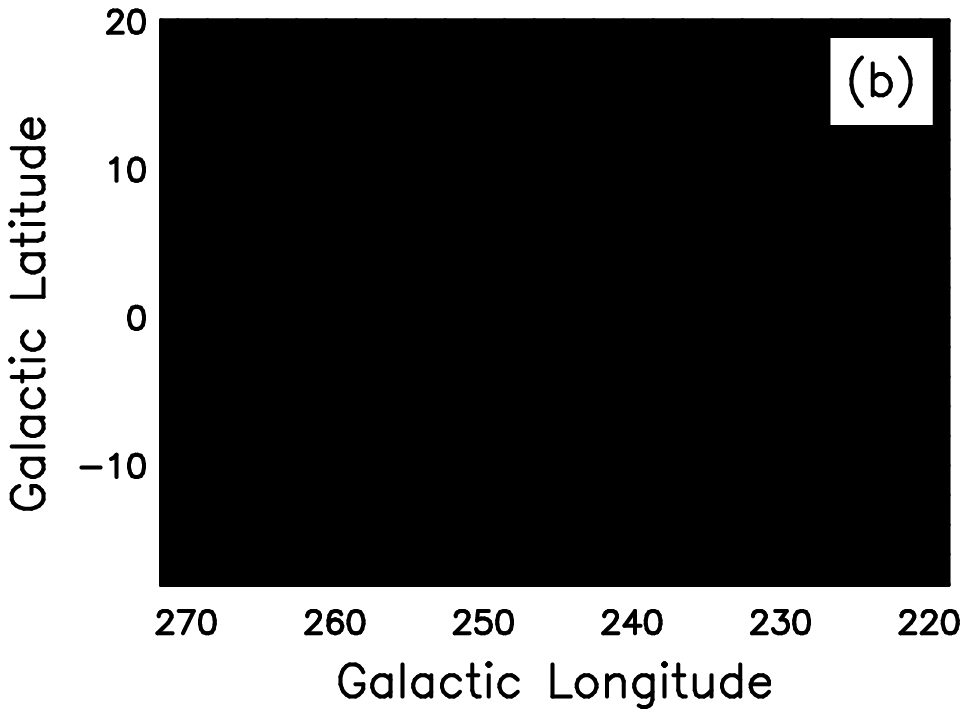}}}
\caption{ (a) Number density of polarised sources in the NVSS
from \cite{stil2007}. Contours are drawn at 2.7 and 4.0 sources per
square degree. The black area in the lower left corner is below the southern
declination limit of the NVSS.  (b) H$\alpha$ intensity in the same
region from the SHASSA survey \cite{gaustad2001}.
\label{Gum-fig}
}
\end{figure}

In addition to probing the magneto-ionic medium of the Milky Way, the
intrinsic polarisation properties of extragalactic sources provide
information on large-scale magnetic fields in distant
galaxies. Observations of galaxies at high redshift will allow us to
probe the cosmic evolution of magnetic fields. The resolution and
sensitivity necessary to image the polarisation of normal galaxies at
high redshift will not be available before the SKA. For SKA
pathfinders, we will be limited to information provided by integrated
polarisation properties.

\section{Integrated polarisation of spiral galaxies}

\begin{figure}
\centerline{\resizebox{11cm}{!}{\includegraphics[angle=-90]{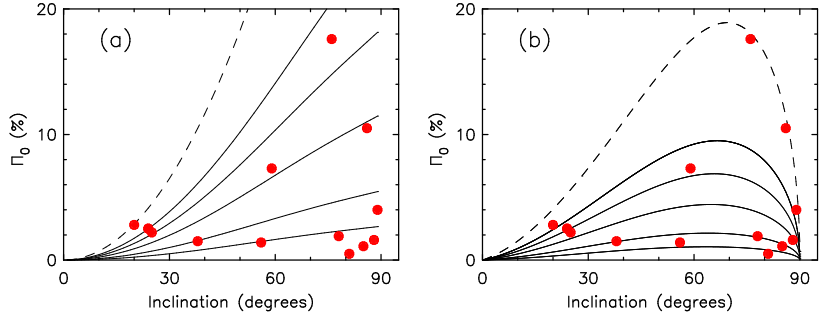}}}
\caption{ Integrated fractional polarisation $\Pi_0$ of nearby spiral
galaxies (dots) with model curves for the expected integrated
polarisation as a function of inclination without Faraday
depolarisation (a) and with Faraday depolarisation (b). The curves
represent models where the intensity of a face-on disk is 5, 10, 20,
30, 40 \% (solid curves from bottom to top), and 70\% (dashed curve)
polarised. 
\label{diskpol-model-fig}
}
\end{figure}

Imaging of synchrotron emission from nearby spiral galaxies shows that
most spiral galaxies have a regular azimuthal magnetic field in the
plane of the disk \cite{beck2005}.  If a spiral galaxy is observed
face-on, the symmetry of the magnetic field implies no direction is
favoured and all polarisation angles are represented equally by
different portions of the disk.  An unresolved axially symmetric
face-on spiral galaxy will thus be an unpolarised radio source. For an
inclined galaxy, the azimuthal magnetic field increasingly projects
along the apparent major axis of the disk. This favours a plane of
polarisation (B rotated by $90^\circ$) aligned with the {\it minor}
axis. An unresolved inclined spiral galaxy is therefore expected to be
polarised with plane of polarisation along the minor axis. This
qualitative conclusion also applies to galaxies with spiral structure.

We investigated the integrated polarised emission from spiral galaxies
by integrating archival Stokes Q and U images over the solid angle of
the disk. Figure~2 shows the resulting integrated $\Pi_0$ at 4.8 GHz
for a sample of 13 nearby galaxies as a function of inclination
\cite{stil2007a}.  The signal to noise ratio of the integrated fluxes
is so high that polarisation bias is negligible.  Some nearby galaxies
would appear as highly polarised radio sources when observed as
unresolved radio sources.  The galaxies with the highest $\Pi_0$ in
Figure~2 are M~31 (17.6\%), NGC~4565 (10.5\%), and M~81 (7.3\%).

Figure~2 also shows curves of predicted integrated $\Pi_0$ based on a
simple axially symmetric model for the polarised intensity in the disk
of a spiral galaxy with and without including the effects of Faraday
depolarisation. Spiral arms, bars and vertical field components will
be included in the future.  The model evaluates magnetic field
components perpendicular to and parallel to the line of sight and from
this the polarised intensity and a simple prescription for Faraday
depolarisation. The curves in Figure~2 form a single-parameter family
where the parameter is the intrinsic fractional polarisation of the
{\it intensity} from the disk as seen face on.

Galaxies at inclination $i \approx 65^\circ$ display the highest
integrated polarisation. The cut off in $\Pi_0$ towards $i = 90^\circ$
results from a strong increase in Faraday rotation because of a longer
pathlength through the disk and larger line of sight components of the
magnetic field, depending on position.  The few galaxies with very
high $\Pi_0$ seem to indicate that the simple treatment of Faraday
depolarisation in our model may overestimate depolarisation effects in
some galaxies. More detailed modeling is in progress.  Approximately
47\% of randomly oriented galaxies will have an inclination between
$50^\circ$ and $80^\circ$, where the model curves predict a broad
maximum. The probability density function of $\Pi_0$ for a randomly
oriented set of spiral galaxies therefore has a maximum near the
highest possible value of $\Pi_0$.

It is interesting to note that the integrated polarisation of spiral
galaxies at low inclination is {\it higher} because of Faraday
depolarisation. At low inclination, Faraday rotation increases mostly
along the apparent major axis of the disk, where the line-of-sight
component of the magnetic field increases linearly with
inclination. The increased Faraday rotation and depolarisation reduce
the axial symmetry of the polarised emission, which is the main source
of depolarisation when integrating over the disk. This effect can be
seen in the model curves of Figure~2 (a) and (b).

Integrated polarisation of spiral galaxies is a promising technique to
study the large-scale magnetic fields in spiral galaxies for large
samples. For an optically selected sample, the available information
is $\Pi_0$ as a function of inclination, and the relation between the
plane of polarisation and the position angle of the minor axis. If
polarimetry at a second wavelength is available for a significant
number of spiral galaxies, e.g. at 1.4 GHz from the GALFACTS survey,
the integrated polarisation of spiral galaxies allows a statistical
study of large-scale asymmetry in the magnetic fields of galaxies, the
occurrence of magnetic fields perpendicular to the disk, and the
amount of Faraday depolarisation as a function of galaxy type.

Barred galaxies have the added complication of a possible magnetic
field oriented along the bar. A sample of barred galaxies
\cite{beck2002} displays similarly high integrated $\Pi_0$
\cite{stil2007a}. Our conclusion that unresolved spiral galaxies can
be highly polarised sources therefore includes barred galaxies.  More
detailed models will include bars and spiral structure in the near
future.

\section{Models of polarised source populations}

The polarised source counts across the transition from AGN to star
forming galaxies can be modeled by convolving total
intensity counts for each population with the appropriate $\Pi_0$
distribution. We apply the $\Pi_0$ distribution of \cite{bg04} to
steep spectrum AGN ($\alpha > 0.5$, $S_\nu \sim \nu^{-\alpha}$), and
the distribution of \cite{tucci2004} to flat spectrum AGN. The $\Pi_0$
distribution for spiral galaxies was derived from the models in
Section~2, and convolved with two model curves for the source counts
of star forming galaxies.  Many luminous starburst galaxies cannot be
adequately represented by a spiral galaxy disk. We take the model
source sounts for star bursts as an upper limit, and the model curve
for a non-evolving local population as a lower limit to the number
counts of star forming galaxies.

Figure~3 (a) shows model total-intensity source counts for different
populations of radio sources, following \cite{hopkins2000}. Model
polarised source counts obtained by convolving each of these curves
with the appropriate $\Pi_0$ distribution are shown in Figure~3 (b),
along with polarised source counts from the DRAO ELAIS N1 deep field
\cite{taylor2007}. The models in Figure~3 do not assume any variation
of fractional polarisation with luminosity or flux density.  The
flatter slope of the normalised polarised source counts for
$p < 4$ mJy seems to be caused by a higher degree of polarisation of
faint steep spectrum AGN.  The excess mJy polarised sources in
Figure~3 (b) were indeed identified with more highly polarised
steep-spectrum AGNs by \cite{taylor2007}.

Why are faint steep-spectrum sources more highly polarised? It is
possible that the large class of steep spectrum radio sources should
be subdivided according to another parameter than spectral index.  A
low-luminosity slowly evolving sub-class of steep-spectrum sources
(see the contribution by E. M. Sadler, this meeting) with a high $\Pi_0$ could
be responsible for the discrepancy between the data and the models.  A
morphological distinction that is related to radio luminosity already
exists in the Fanaroff \& Riley classification \cite{FR}, but there is
no evidence at this time that the faint polarised sources are
indeed distant FR I sources.

\begin{figure}
\resizebox{7.5cm}{!}{\includegraphics[angle=0]{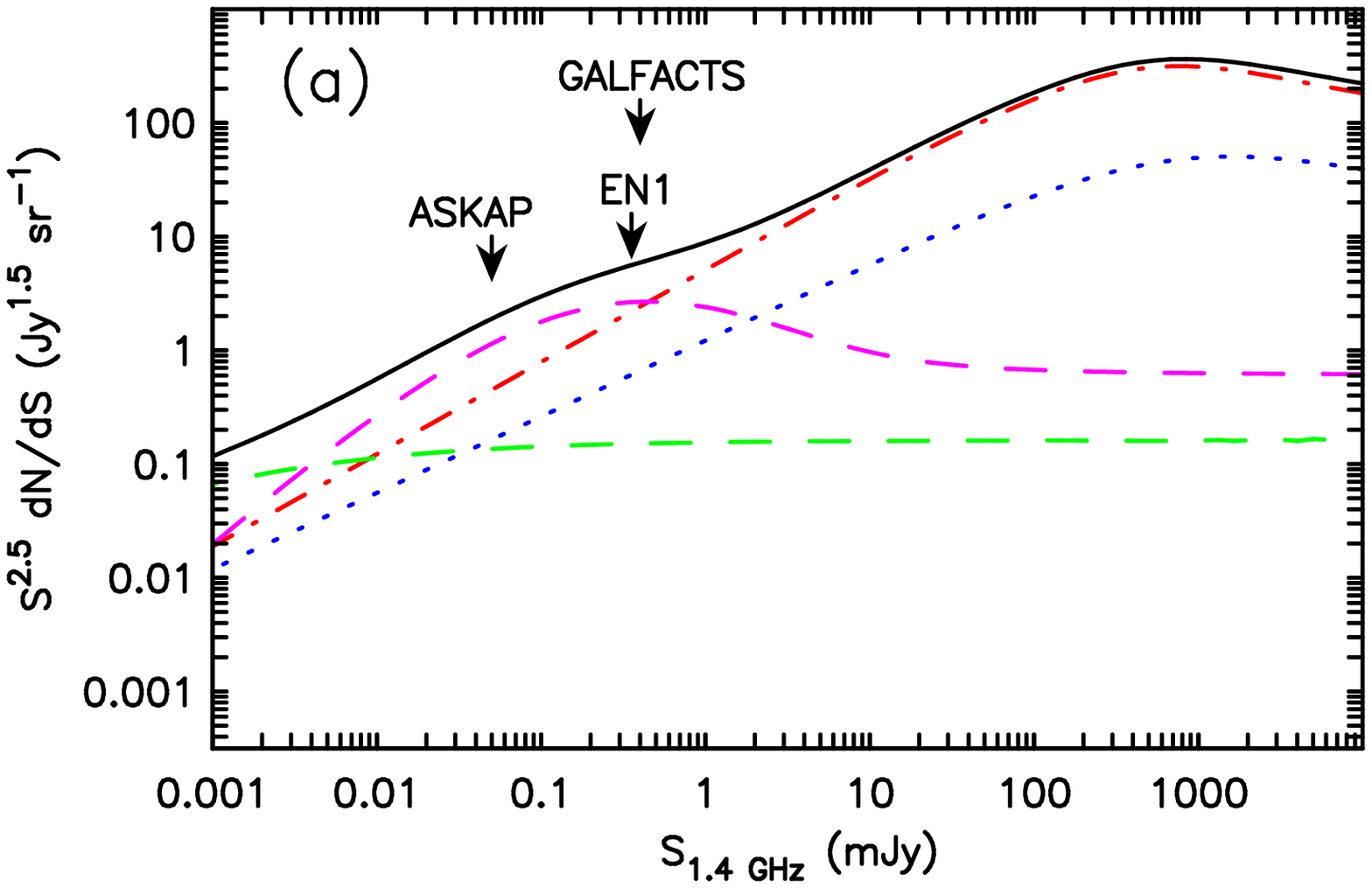}}
\resizebox{7.5cm}{!}{\includegraphics[angle=0]{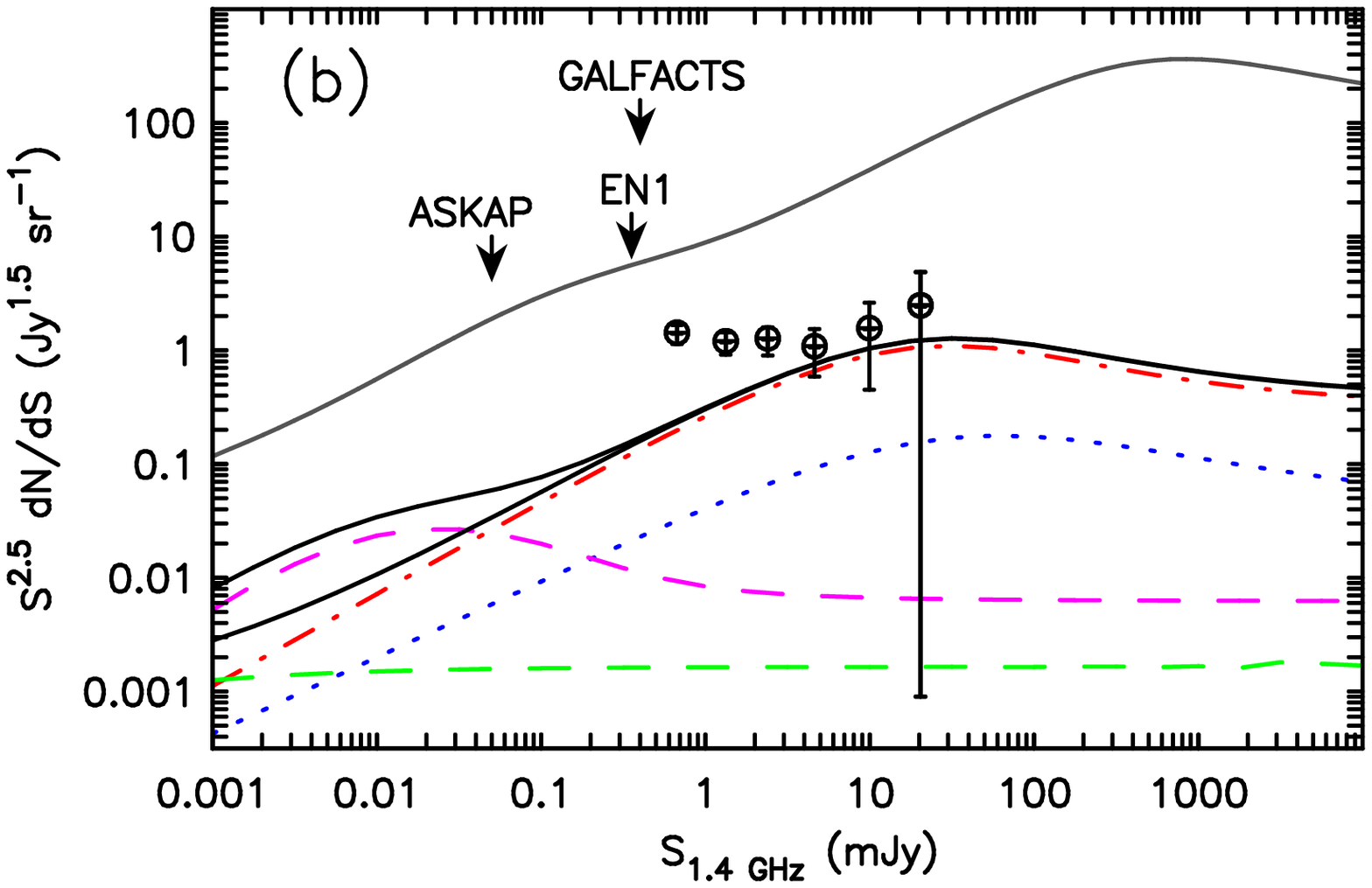}}
\caption{ (a) Radio source counts calculated following
\cite{hopkins2000}. Steep spectrum AGN: red dot-dashed curve. Flat
spectrum AGN: blue dotted curve. Starburst galaxies: magenta dashed
curve. Non-evolving spiral galaxies: green dashed curve. Black curve:
total source counts. (b) Model polarised source counts obtained by
convolving the curves in (a) with the $\Pi_0$ distributions described
in the text. The data are observed polarised source counts from
\cite{taylor2007}. The $5\sigma$ point source sensitivity of three
polarisation surveys is indicated.
\label{modelcounts-fig}
}
\end{figure}

\end{document}